\documentclass[conference]{IEEEtran}
\IEEEoverridecommandlockouts
\usepackage{cite}
\usepackage{amsmath,amssymb,amsfonts}
\usepackage{algorithmic}
\usepackage{graphicx}
\usepackage{textcomp}
\usepackage{balance}
\usepackage{xcolor}
\usepackage{xspace}
\usepackage[hyphens]{url}
\usepackage{hyperref}
\usepackage{tcolorbox}
\usepackage{tabularx}
\usepackage[normalem]{ulem}
\usepackage{listings}
\usepackage{enumitem}

\def\BibTeX{{\rm B\kern-.05em{\sc i\kern-.025em b}\kern-.08em
    T\kern-.1667em\lower.7ex\hbox{E}\kern-.125emX}}
\begin{document}

\def\correspondingauthor{\footnote{Corresponding author.}}

\newcommand{\method}{{\textsc{Picaso}}\xspace}
\newcommand{\so}{{Stack Overflow}\xspace}

\newboolean{showcomments}
\setboolean{showcomments}{true}
\ifthenelse{\boolean{showcomments}}
 { \newcommand{\mynote}[2]{
      \fbox{\bfseries\sffamily\scriptsize#1}
        {\small$\blacktriangleright$\textsf{\emph{#2}}$\blacktriangleleft$}}}
        { \newcommand{\mynote}[2]{}}
\newcommand{\todoc}[2]{{\textcolor{#1} {\textbf{#2}}}}
\newcommand{\todo}[1]{{\todoc{red}{\textbf{#1}}}}
\newcommand{\todored}[1]{\todoc{red}  {\textbf{#1}}}
\newcommand{\todoblue}[1]{\todoc{magenta}  {\textbf{#1}}}
\newcommand{\todopp}[1]{\todoc{blue}  {\textbf{#1}}}

\newcommand{\ft}[1]{\mynote{Ferdian}{\todored{#1}}}
\newcommand{\kisub}[1]{\mynote{Kisub}{\todored{#1}}}
\newcommand{\zt}[1]{\mynote{Ting}{\todopp{#1}}}
\newcommand{\iv}[1]{\mynote{Ivana}{\todoblue{#1}}}

\setlist{noitemsep} 
\lstset{
	columns=fullflexible,
	,basicstyle=\footnotesize\ttfamily
	,stringstyle=\footnotesize\ttfamily
	,aboveskip={1pt}
	,belowskip={1pt}
	,showstringspaces=false 
	,numberstyle=\tiny
        ,breaklines=true
	,tabsize=3
	,escapeinside={(@}{@)}
}

\pagestyle{plain}

\title{\method : Enhancing API Recommendations with Relevant Stack Overflow Posts}

\author{\IEEEauthorblockN{Ivana Clairine Irsan, Ting Zhang*\thanks{*Corresponding author.}, Ferdian Thung, Kisub Kim, and David Lo}
\IEEEauthorblockA{School of Computing and Information Systems, Singapore Management University\\
Email: \{ivanairsan,\:tingzhang.2019,\:ferdianthung,\:kisubkim,\:davidlo\}@smu.edu.sg}
}

\maketitle

\begin{abstract}

While having options could be liberating, too many
options could lead to the sub-optimal solution being chosen.
This is not an exception in the software engineering domain. Nowadays,
API has become imperative in making software developers’
life easier. APIs help developers implement a function 
faster and more efficiently. However, given the large number
of open-source libraries to choose from, choosing the right APIs is not a simple task. Previous studies on API recommendation leverage natural language (query) to identify which API would be
suitable for the given task. However, these studies only consider
one source of input, i.e., GitHub or Stack Overflow, independently. 
There are no existing approaches that utilize
Stack Overflow to help generate better API sequence recommendations
from queries obtained from GitHub. Therefore, in this study, we
aim to provide a framework that could improve the result of the
API sequence recommendation by leveraging information from
\so. In this work, we propose \method, which leverages a bi-encoder to do contrastive learning and a cross-encoder to build
a classification model in order to find a semantically similar Stack Overflow post given an annotation (i.e., code comment).
Subsequently, \method then uses the Stack Overflow’s title as a query expansion. \method then
uses the extended queries to fine-tune a CodeBERT, resulting in an
API sequence generation model. Based on our experiments, we
found that incorporating the Stack Overflow information into CodeBERT
would improve the performance of API sequence generation's BLEU-4 score by 10.8\%.

\end{abstract}


\section{Introduction}
\label{sec:intro}
In modern software development, Application Programming Interfaces (APIs) and libraries have been widely used by developers to improve development efficiency.
Considering the large volume of existing APIs, multiple APIs can be used to implement the same functionality.
For instance, to read a file in Java using the JDK library, a developer could use either FileReader, BufferedReader, Files, Scanner, etc., depending on the size of the file and type of the data that need to be read (i.e., byte or characters).
While having many options is liberating, the problem of choosing the right API for a certain functionality arises.
Given the enormous number of available APIs, it is challenging for developers to find and learn which API to use along with the complementary APIs needed in the \textcolor{black}{API invocation} sequence~\cite{peng2022revisiting}.
Therefore, it is desirable to have an API recommendation system that can recommend the correct APIs to use~\cite{deepapi,fowkes2016parameter}.

To support developers in finding the correct APIs, several API recommendation methods have been proposed~\cite{huang2018api,rahman2016rack,wei2022clear,chen2021holistic}.
Depending on the input type, these methods can be categorized into query-based or code-based API recommendation~\cite{peng2022revisiting}.
In this work, we mainly focus on the former type of API recommendation. 
We are concerned with how to accurately translate a natural language query that describes the intention of the programming task to an API sequence that implements the requirements.
While many query-based API recommendation approaches have been proposed~\cite{wei2022clear,huang2018api,rahman2016rack}, only a limited number of them recommend API sequence~\cite{deepapi,martin2022deep}. 
The best of which is the work by Martin and Guo~\cite{martin2022deep} that leverages CodeBERT, a bimodal pre-trained model for programming language and natural language, to generate API sequences. 

In practice, to search for suitable APIs, developers tend to resort to \so~\cite{huang2018api}.
Prior studies~\cite{wei2022clear,huang2018api,rahman2016rack} have shown that \so can be effectively utilized to recommend a set of APIs as it contains tremendous crowd-knowledge.  
Additionally, diversifying the resources for \textcolor{black}{training an API recommendation model} has been investigated by researchers~\cite{wang2018entagrec++,wei2022clear,ponzanelli2014prompter, zhou2021boosting} and verified to be beneficial.

Previous studies\cite{huang2018api,wei2022clear} utilize \so data by finding similar \so posts and recommending curated APIs mentioned in the answers.
Their objective is to correctly identify the similar \so posts for a question that a developer comes up with.
However, developers' need for good API recommendations goes beyond a question when they face difficulties.
On a daily basis, they implement functions to perform certain tasks that could be summarized in the form of natural language (NL), i.e., code comment/annotation.
Oftentimes, the NL representation of the code could be paired with a \so post (example presented in Section \ref{sec:motivating-example}). This motivates us to explore the potential of leveraging \so information in search of generating better API sequence recommendations.



\textcolor{black}{
With a goal to ease developers' daily tasks, we propose \method (Enhancing A\underline{PI} Re\underline{c}ommend\underline{a}tions with Relevant \underline{S}tack \underline{O}verflow Posts), an API recommendation technique that leverages multi-source (i.e., \so and GitHub) information via query expansion.
\method takes a user query as either code comment or annotation, then finds the \so post that is most similar to the query. 
Once it discovers a similar post, the post is delivered to the CodeBERT encoder as query expansion to generate the API sequence that implements the functionality described in the query.
To find the most similar \so post for a given annotation, we adapted a framework \textcolor{black}{proposed by Wei et al.}~\cite{wei2022clear}, who propose a state-of-the-art technique in finding similar \so posts for a \so query. The framework consists of a filtering and a re-ranking model.
The filtering model is built on top of the contrastive learning method, while the re-ranking model is trained by performing joint-embedding training.}

We evaluated \method on a dataset derived from the work of Martin and Guo~\cite{martin2022deep}. \textcolor{black}{\method manages to outperform the state-of-the-art API sequence recommendation model by 10.8\% in terms of BLEU-4 score.}

Our contributions can be summarized as follows:
\textcolor{black}{
\begin{itemize}
    \item{We are the first to show that \so posts can boost the effectiveness of the query-based API sequence recommendation, particularly a query expansion.}
    \item{We propose \method, a multi-source API sequence recommendation method that can outperform the state-of-the-art approach.}
    \item{We conduct experiments to demonstrate the effectiveness of \method, achieving a \textcolor{black}{10.8\% improvement measured in BLEU-4 score.}}
\end{itemize}
}

The remainder of the paper is organized as follows. 
Section~\ref{sec:background} introduces the problem formulation and a motivating example. 
We describe our approach in Section~\ref{sec:approach}. 
Section~\ref{sec:setting} \textcolor{black}{presents} our experimental setup and the research questions. 
Experimental results are \textcolor{black}{described} in Section~\ref{sec:result}. 
We discuss the results in Section~\ref{sec:discussion}. 
Section~\ref{sec:related} \textcolor{black}{presents} the related works. 
We conclude our work \textcolor{black}{and present} future work in Section~\ref{sec:conclusion}.






\section{Preliminaries}
\label{sec:background}

\subsection{Problem Formulation}
\label{sec:apirec}
Following prior works~\cite{deepapi,martin2022deep}, we formulate the API recommendation task as a sequence-to-sequence task.
Given a query in the form of \textcolor{black}{annotation, i.e., code comment that describes the purpose of a function}, our goal is to recommend a sequence of API calls that implements the functionality described in the query.
We also aim to leverage information from multiple sources such as \so and GitHub.

Here, we formally define the task of query-based API recommendation.
We denote the input sequence, e.g., annotation, as $W = \langle W_1, W_2, \ldots, W_n\rangle$, where $W_i = \langle w_i^1, w_i^2, \ldots, w_i^{n_i}\rangle$ and $w_i^j$ refers to the $j^{th}$ subtoken in word $W_i$.
We want to generate an API sequence to be recommended.
We denote the API sequence as $A = \langle A_1, A_2, \ldots, A_m\rangle$, where $A_i = \langle s_i^1, s_i^2, \ldots, s_i^{n_i}\rangle$.
$s_i^j$ refers to the $j^{th}$ subtoken in API call $A_i$.

When a model predicts the subtoken $s_i^1$, the input of the model would be the input sequence $\langle W_1, W_2, \ldots, W_n\rangle$ and the previously predicted subtokens $\langle s_1^1, \ldots, s_1^{n_1}, \ldots, s_{i-1}^{1}, \ldots, s_{i-1}^{n_{i-1}}\rangle$. 
When the end token is predicted, an API recommendation model will output the chain of predicted subtokens as the API sequence.
Take the annotation and the API sequence in the first box in Figure~\ref{fig:rel-post} as an example.
In the beginning, the input is only the annotation.
After the API recommendation model predicts two subtokens in the target API sequence, the input will be $\langle W_1, W_2, W_3, W_4, s_5^1, s_5^2\rangle$, where $W_1=parse$, $W_2=string$, $W_3=to$, $W_4=object$, $s_5^1=Float$, $s_5^2=.$.
The API recommendation model would predict the next token based on this input.
Suppose the model predicts $parseFloat$ and $\langle$\texttt{eos}$\rangle$, the output API sequence would be the combination of these three subtokens $s_5^1,s_5^2$ and $parseFloat$.

\subsection{Motivating Example}
\label{sec:motivating-example}
\begin{figure}
\centering
\includegraphics[width=\linewidth]{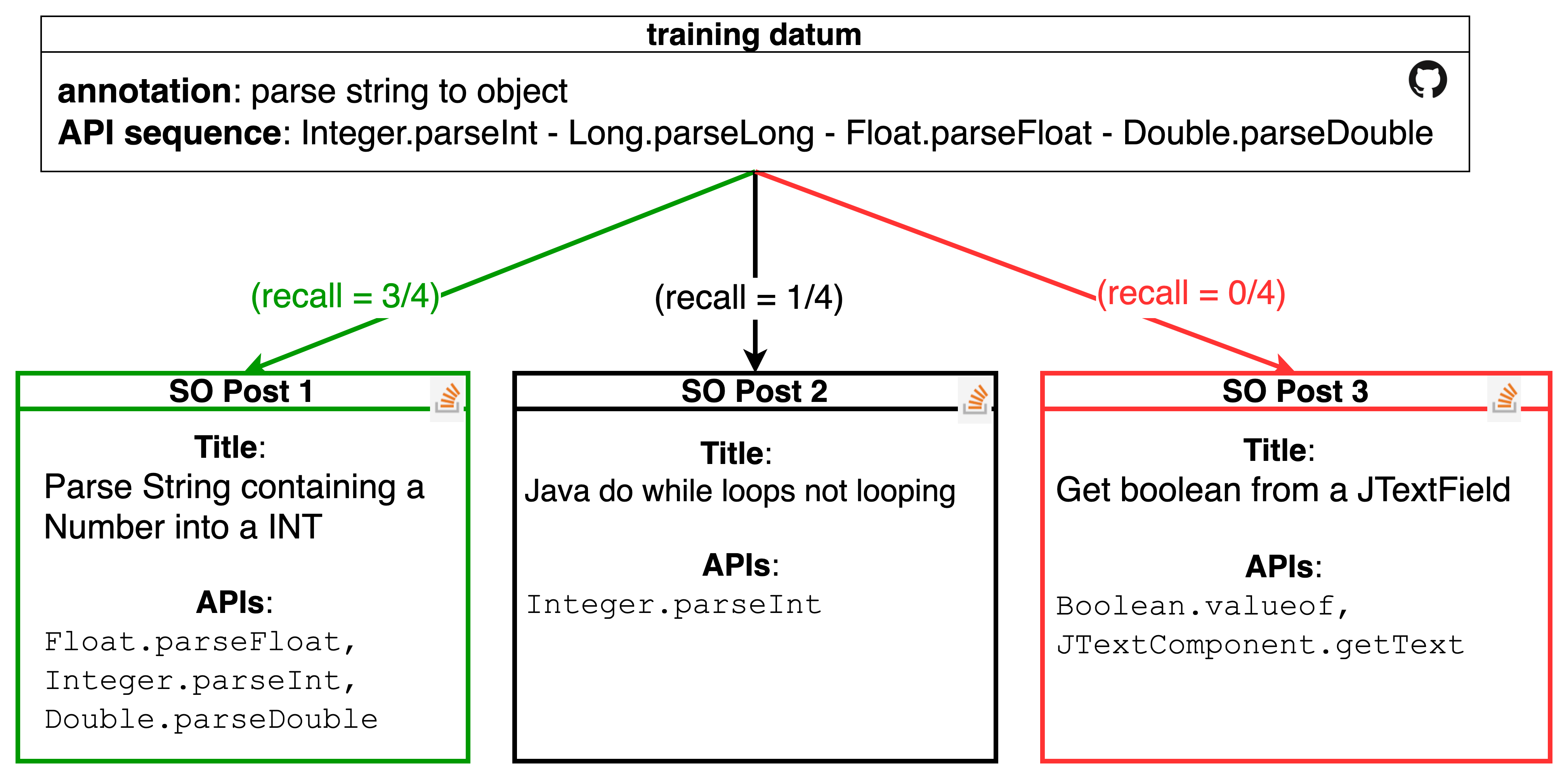}
    \caption{Finding relevant \so post based on API in training data and API mentioned in the SO post's answer}
    \label{fig:rel-post}
\end{figure}

We use the example in Figure~\ref{fig:rel-post} to motivate the need for leveraging \so posts for API sequence recommendation. The annotation for this example is \texttt{parse string to object}, and the target APIs are \texttt{Integer.parseInt}, \texttt{Long.parseLong}, \texttt{Float.parseFloat}, and \texttt{Double.parseDouble}. From the annotation alone, it is not possible to know that the input \texttt{String} contains numbers, and the \texttt{Object} should be an instance of numbers (e.g., \texttt{Integer}). 
Broadly speaking, the input \texttt{String} can contain any set of characters, and the output \texttt{Object} can be an instance of any object.

To obtain a more specific query, we can expand based on the \so posts that may contain similar sets of APIs. 
\textcolor{black}{We define the similarity between an annotation and a \so post as the proportion of APIs from the target APIs that are mentioned in the post. The more APIs from the target APIs that are mentioned in the \so post, the more similar the \so post is to the annotation.
Based on this definition, we can prepare the training data which contains the most similar \so post and the annotation; and learn a similarity function from it.
In practice, in which the target APIs are not available, 
we can use the learned similarity function to find the most similar posts given the annotation. 
We describe in detail how we construct such a function in Section~\ref{sec:similarso}.
}
 
Consider that we have a collection of \so posts containing SO Post 1, SO Post 2, and SO Post 3 as shown in Figure~\ref{fig:rel-post}.
Among these \so posts, SO Post 1 would be the most similar since it contains the largest number of relevant APIs: \texttt{Float.parseFloat}, \texttt{Integer.parseInt}, and \texttt{Double.parseDouble}.
The title for SO Post 1 is \texttt{parse String containing a Number into a INT}. 
By expanding the query with this title, we can now understand that the input \texttt{String} should be the numbers.

The above illustration demonstrates that we can leverage Stack Overflow to improve the API sequence recommendation, particularly by adding more information from the \so so that we can input a more accurate query to the API sequence generation model.

\section{Approach}
\label{sec:approach}

\begin{figure}
\centering
\includegraphics[width=0.5\textwidth]{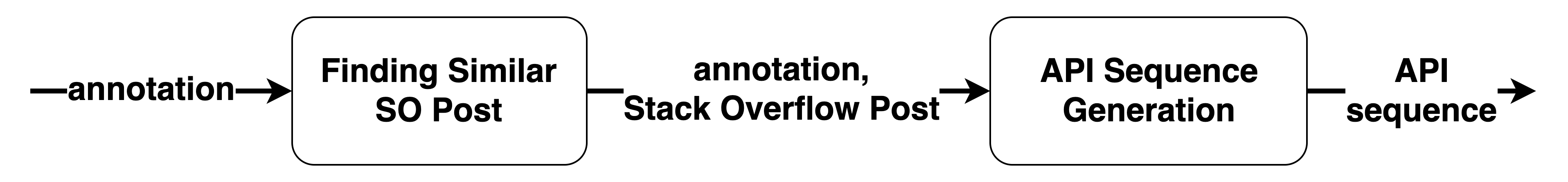}
    \caption{\method high-level architecture}
    \label{fig:overall-architecture}
\end{figure}

\begin{figure*}
\centering
\includegraphics[width=0.95\textwidth]{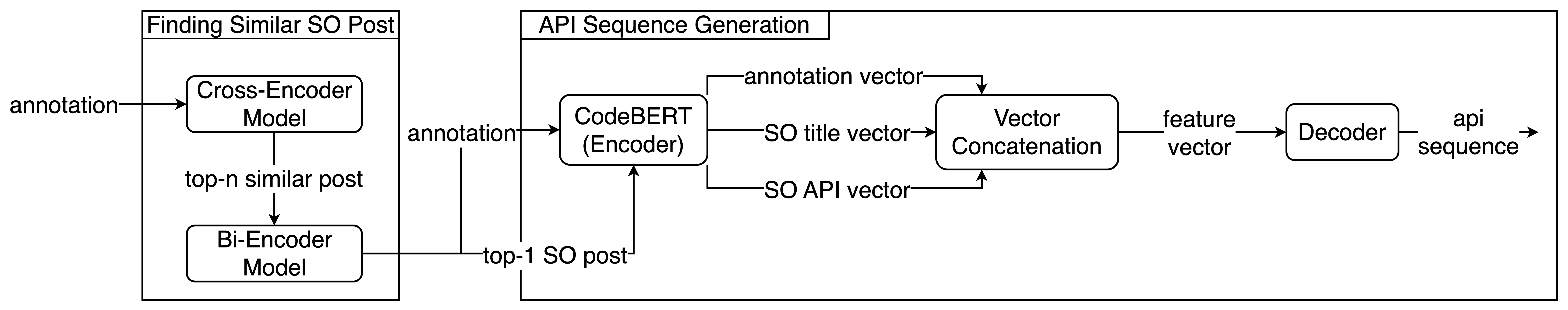}
    \caption{\method detailed architecture}
    \label{fig:model}
\end{figure*}

In this section, we introduce the overall architecture of \method and discuss the details of each component.
\textcolor{black}{The overall architecture is presented in Figure~\ref{fig:overall-architecture}.}

\textcolor{black}{In general, 
\method consist of two phases: 1) \textbf{Finding Similar SO Post}. 
This phase is used to find \so post that is the most semantically similar to the input annotation (i.e., code comment) and 
2) \textbf{API sequence generation}.
This takes annotation and the \so information retrieved from the first phase to generate an API sequence.}

\textcolor{black}{For the Finding Similar SO Post phase, we adapt CLEAR's~\cite{wei2022clear} framework, which has been modified to leverage the sentence embedding technique for finding  relevant posts.} 
While CLEAR leverages the sentence embeddings to map a \so post towards other similar \so posts, we utilize sentence embedding to map an annotation towards similar \so posts.
Therefore, we leverage the usage of contrastive learning and joint embedding training while implementing our adaptation of the triplet data generation.
For the second phase, we leverage CodeBERT as the base encoder for the encoder-decoder architecture since it is state-of-the-art for API sequence generation technique~\cite{martin2022deep}.
Further explanation is given in Section~\ref{sec:apigeneration}.

\textcolor{black}{
The detailed architecture of \method is presented in Figure~\ref{fig:model}. It shows the choices of the model used in every component of \method, as well as the data flow.
}


\begin{figure}
\centering
\includegraphics[width=\linewidth]{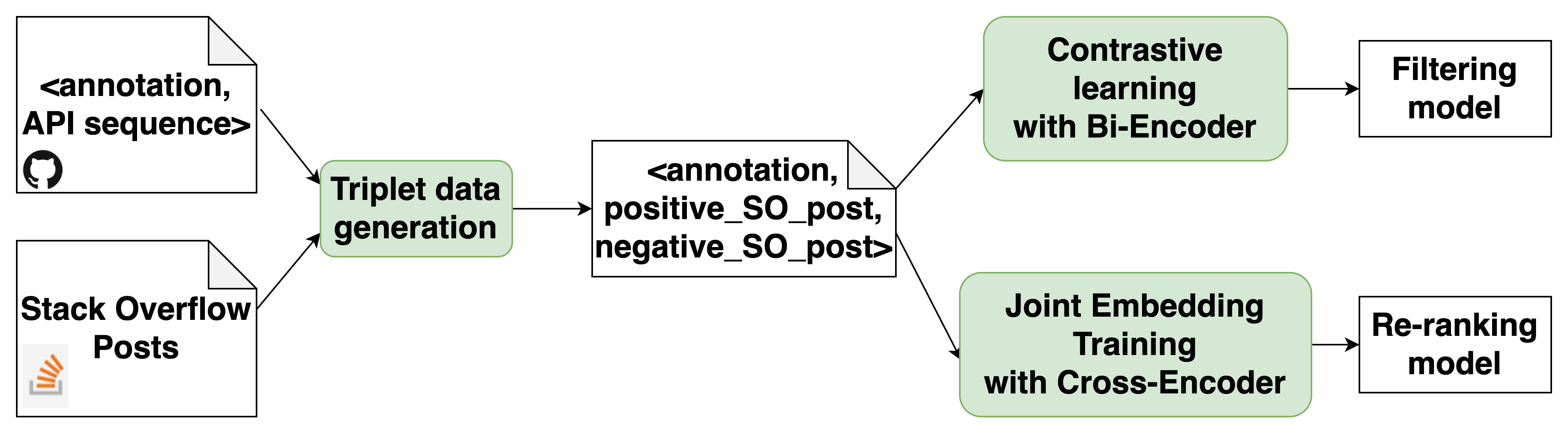}
    \caption{Training process for finding the most similar \so post}
    \label{fig:clear}
\end{figure}

\subsection{Finding Similar SO Post}

\label{sec:similarso}
To find the most similar \so post, we leverage contrastive learning~\cite{oord2018representation} and joint embedding training~\cite{devlin2018bert}.
\textcolor{black}{The same approach has been used in CLEAR\cite{wei2022clear}, with a difference in domain of the data.
The objective of CLEAR is to find the most relevant \so posts for a given \so post question.
In other words, they aim to build a better sentence representation for two natural language texts from the same domain.}
On the contrary, in this work, we aim to train a sentence embedding model that could place an annotation derived from GitHub's project towards \so post, which comes from different domains.


Although both GitHub and \so are from the software engineering domain, the nature of the content is essentially different.
For example, the nature of the natural language (NL) from \so comes in the form of a question, oftentimes related to an error rather than a technical functionality of an API.
Moreover, the APIs mentioned in the answer posts are not necessarily a sequence.
On the other hand, annotation in GitHub data portrays the functionality of a task in the form of a statement rather than a question.
Moreover, the APIs extracted from GitHub are presented in the form of a sequence.
\textcolor{black}{We believe that adapting contrastive learning and joint embedding training in our approach as an attempt to bring similar annotation and \so post together in the sentence embedding is advantageous.
Sequentially, the most similar post retrieved in this phase will be used for query expansion in the API sequence generation phase.}

The architecture of the framework to find the most relevant post is presented in Figure~\ref{fig:clear}.
\textcolor{black}{The filtering and re-ranking model produced in this step will be utilized to find the most semantically similar post given an annotation.
Finding a semantically relevant post is imperative, as we utilize such a post as query expansion that will directly affect the performance of our API generation model.}
The details for each process are explained as below:


 \textcolor{black}{\textbf{Triplet Data Generation:}
In order to train both the filtering model via contrastive learning and the re-ranking model via joint embedding, we need to prepare the training and validation data that are built from the training set of our dataset.
The training process takes triplet data as input containing an annotation, a positive, and a negative \so title.
To accommodate the triplet data generation, we need to find highly similar annotations and \so pair to be used as positive pair.
Following the motivation in Section~\ref{sec:motivating-example}, we determined the relevancy of a post and annotation based on the \textcolor{black}{{\em overlap rate}}
between the APIs appearing in the API sequence from GitHub (which corresponds to the annotation) and the APIs mentioned in the \so answer.}
Figure~\ref{fig:rel-post} shows an example of similar posts, given an annotation and its corresponding API sequence.
We first convert the API sequence in the training datum into a set of APIs, which we refer to as \textit{training API set}.
Towards this direction, we calculate the overlap rate of the APIs mentioned in the answer post (i.e., the proportion of APIs in the training API set).
In the example, the training datum is considered similar to the SO Post 1\footnote{https://stackoverflow.com/questions/22016489/parse-string-containing-a-number-into-a-int} because 3 out of 4 in the training API set are mentioned in the answer post.
Therefore, SO Post 1 is considered a positive training post for contrastive learning.
On the other hand, SO Post 3\footnote{https://stackoverflow.com/questions/8706070/get-boolean-from-a-jtextfield} is considered a negative post as it does not contain APIs from the API training set.
For this work, we use $0.75$ as the threshold for \textcolor{black}{the overlap rate as we consider this threshold captures high-quality data points while ensuring we have enough data points to train the Bi-Encoder (i.e., via contrastive learning) and Cross-Encoder (i.e., via joint-embedding training).}

Due to this threshold, for SO Post 2\footnote{https://stackoverflow.com/questions/26108503/java-do-while-loops-not-looping}, even though it has one matching API, it is not counted as the positive example for the training datum since the overlap rate is only 0.25, which is less than the $0.75$ threshold.


Once we discover all of the positive posts for each training datum, we randomly choose five posts with a really small \textcolor{black}{overlap rate} (i.e., less than $0.01$) and use them as negative posts.
Pairs of positive and negative posts are then formed into triplets to be used in contrastive learning.
Note that some training data are naturally discarded because they do not have a positive \so post with an overlap rate above the threshold.
Following a prior study~\cite{wei2022clear}, we used two parameters, namely \textit{p} and \textit{n} to build the triplets, where \textit{p} refers to the number of positive examples and \textit{n} refers to the number of negative examples. 
We produced \textit{p} x \textit{n} numbers of triplets for each annotation \textit{A}.
Based on their grid search experiment, using \textit{p}=10 and \textit{n}=10 performs the best.
Following this, ideally, each annotation should produce 100 triplets for contrastive learning. However, this is not the case for every annotation.
If \textit{A} only has three positive posts to be paired with, it will only have 30 triplets in the end.
If \textit{A} has more than ten positive posts, we randomly choose ten of them to generate 100 triplets for annotation \textit{A}.

\textbf{Contrastive Learning}: 
\textcolor{black}{
We leverage contrastive learning~\cite{oord2018representation} as the first step to learn the semantic relationship between annotation and \so post.
The contrastive training is done by fine-tuning a RoBERTa~\cite{liu2019roberta} based Cross-Encoder~\cite{reimers2019sentence} model to learn a better semantic embedding representation. 
Even if an annotation and \so posts do not share mutual tokens lexically, the sentence embedding should be able to put the annotation and related \so posts in a relatively close vector space if they share the same topic semantically.
The learning goal is to obtain a representative sentence embedding that could serve as a filtering model. 
It should be able to effectively fetch the top-n most similar \so posts by fetching \so posts whose embedding has the highest cosine similarity score with the annotation embedding.
}


\textbf{Joint Embedding Training}: To train a BERT as a classification model, joint embedding training \cite{devlin2018bert} is commonly used in practice.
In this work, we employ a RoBERTa-based Cross-Encoder\footnote{https://huggingface.co/cross-encoder} model for the classification task.
The purpose of the classification is to re-rank the filtered posts returned in the previous step.
The training data for classification is obtained from the triplet generated for contrastive learning.
We transform the triplets to make them compatible with the classification task.
For each triplet, we assigned the annotation and its positive post with label {\tt 1} and we assigned label {\tt 0} for the negative post pair.
The result of this joint embedding training is a classification model that can be used to perform a re-ranking for the filtered post.
We leverage the probability score as a measure to determine the rank of the relevant post and take the top-1 post to be used in the API sequence generation.

\textcolor{black}{
Finally, after the contrastive learning and joint embedding training are done, the models produced by these learning processes will be used to pair an annotation and its target APIs with the most relevant \so post. Using the trained cross-encoder model, we produce embeddings for the annotation and \so posts' titles. We take the top-10 most relevant \so posts in terms of their cosine similarity between the annotation and \so title embedding. We then passed the top 10 posts to the bi-encoder model to be re-ranked based on the probability score. We regard the top-1 post after the re-ranking process as the most semantically relevant post for the given annotation.}


\subsection{API Sequence Generation}
\label{sec:apigeneration}
\method leverages Cross-Encoder and Bi-Encoder models trained in the previous phase to build the training data used in API sequence generation.
The training data is then used to fine-tune the CodeBERT encoder-decoder model to generate an API sequence.
Before the API sequence generation training phase, we leveraged the trained Cross-Encoder and Bi-Encoder to obtain the \so post that is most similar to the annotation.
Given the pair of an annotation and its most similar \so post, we extract the title and \so API from the post and build a dataset that consists of \textit{annotation, \so title}, \textit{\so API set}, and \textit{target API sequence}.
To generate the API sequence, we adopt the encoder-decoder framework.

\textbf{Encoder:} The encoder is used to map the input sequence into a continuous vector representation.
We encode the annotation and the \so components (i.e., \so title and API) with a Transformer-based encoder separately.
We initialize the parameters of each encoder with the pre-trained CodeBERT model.
Secondly, we concatenate the individual encoded vectors (i.e., annotation vector, SO title vector, and SO API vector) into a single vector referred to as the feature vector.
Last, the feature vector is passed through a Transformer decoder with 6 layers, where the parameters are randomly initialized.

\textbf{Decoder:} The decoder is used to generate the API sequence. 
The decoding starts with a start token.
At each position, the decoder predicts the target token by taking the encoder output and previously generated output token list as input.
The decoder stops decoding when it generates the end token.

\textbf{Beam Search.} During the decoding stage, we aim to generate an accurate API sequence.
However, it is hard to guarantee the correctness of each subtoken at each prediction position.
If the model makes a wrong prediction at a certain position, it will affect the following predictions.
Thus, instead of considering only a single subtoken every time, one can consider all the subtokens in each position.
The obvious drawback is the explosion of all the combinations.
To achieve the balance between effectiveness and efficiency, we adopt the widely-used beam search~\cite{freitag-al-onaizan-2017-beam,shu-nakayama-2018-improving} to search for the recommended API sequence heuristically.
Simply put, in each prediction, beam search considers the top-k subtokens (k refers to the size of the beam) based on conditional probability.
Following the existing works~\cite{freitag-al-onaizan-2017-beam,kang2021apirecx}, we adopt the Equation~\ref{equ:cond_prob} to calculate the chain probability of a chain of the subtokens: 

\begin{equation}
    \label{equ:cond_prob}
    P\left(s_m^1, \ldots, s_m^i \mid w_1^1, \ldots, w_1^{n_1}, \ldots, w_{m-1}^{n_{m-1}}\right)=\prod_{j=1}^i p\left(s_m^j\right)
\end{equation}

\noindent where $p(s_m^j)$ refers to $p(s_m^j \mid w_1^1, \ldots, w_1^{n_1}, \ldots, w_{m-1}^{n_{m-1}}, s_m^1, \ldots, s_m^{j-1})$).
It represents the probability of the $j^{th}$ subtoken in the chain of $(s_m^1, \ldots, s_m^i)$.

\section{Experimental Setting}
\label{sec:setting}
In this section, we elaborate on our experiment setup, including the dataset, baseline approach, and evaluation metrics.
Next, we present our experimental results that answer a few research questions.

\subsection{Dataset}
We use the dataset provided by Martin and Guo \cite{martin2022deep}, which is made of pairs of annotation and API sequences.
It is derived from the DeepAPI dataset \cite{deepapi} and was cleaned by removing duplicates.
The DeepAPI dataset contains more than 7 million pairs of annotation and API sequences for training and 10,000 pairs for testing.
However, Martin and Guo found that there are duplicates in the dataset.
There were also some pairs in the test that appear in the training.
After removing the duplicates, there are 1,880,472 training pairs and 2,441 test pairs remained.
As for the Stack Overflow dataset, we utilized the dataset used by CLEAR~\cite{wei2022clear}.
This dataset originated from BIKER~\cite{huang2018api}, containing Stack Overflow posts that are related to Java JDK programming topic, have a positive score, and have at least one accepted answer with API entities mentioned in it.

In this paper, we built our dataset by using the deduplicated DeepAPI dataset as the starting point.
We combined both 1,880,472 training pairs and 2,441 test pairs and further processed them as follows:
\begin{itemize}
\item{\textbf{\so API selection}}

In this phase, we gathered APIs mentioned in the \so posts answer and counted their frequency. Following that, we filtered out \so API that is mentioned less than five times in the entire \so posts. This ensures that the APIs appear at least 5 times, which should provide enough samples for the model to learn properly.
In total, we obtained 1,398 API methods that constitute the \textit{API vocabulary} in our dataset. 

\item{\textbf{Dataset filtering}} 

Based on the API vocabulary constructed in the previous step, we filtered out pairs of annotation and API sequences. We removed pairs whose target API contains API(s) that are not within the API vocabulary.
This action ensures that the pairs share the same vocabulary with the \so dataset.
Thus, we can evaluate whether considering multiple sources of information helps (or not) when both sources of information contain information about specific APIs.
\end{itemize}

In total, we collected 196,276 pairs of annotation and API sequences in our dataset.
We then split the dataset into the train, validation, and test sets with a ratio of 8:1:1.
All in all, we have 157,020 pairs in the training set, 19.628 pairs in the validation set, and 19,628 pairs in the test set.

\subsection{Baseline}
\label{sec:baseline}

Based on the previous work conducted by Martin and Guo~\cite{martin2022deep}, CodeBERT is the state-of-the-art approach for generating API sequences.
It surpassed DeepAPI~\cite{deepapi}, the former state-of-the-art in the domain of API sequence generation.
Therefore, we chose to adapt CodeBERT as our baseline model. All experiments were done with the same CodeBERT structure, which is adapted from CodeBERT project~\cite{feng2020codebert}.
CodeBERT is a pre-trained model built specifically for solving tasks related to programming languages.
It is trained on 6 different natural language and programming language pairs, resulting in a powerful pre-trained model whose embedding is proven useful for varying downstream tasks~\cite{zhou2021assessing, mashhadi2021applying}.

\subsection{Hyperparameter Setting}
\label{sec:hyperparameter}
We used the same hyperparameters settings in all of our experiments.
We set the maximum token length of the input (i.e., annotation) and the target sequence to be 64.
For the CodeBERT model that involves multiple inputs (e.g., \so title and API), we also set the maximum token length of 64 for each input.
All models were trained for 30 epochs.

\subsection{Evaluation}
\label{sec:evaluation}
In order to measure the performance of API sequence generation, we employ BLEU score\cite{papineni2002bleu} as the selected metric.
We compare the BLEU score of \method and the baseline to identify which one is the better approach in generating API sequence recommendations.
BLEU-4 score is capable of gauging how accurate a sequence that is generated by a model is compared to the correct target sequence (i.e., ground truth).
As a metric that is widely adopted in machine translation problems, BLEU score is relevant to be used in comparing the automatically generated API sequence against the human-written API sequence and has been used in existing work in API sequence recommendation~\cite{deepapi, martin2022deep, luong2015effective}.

\noindent BLEU score is expressed mathematically as below:
\begin{equation}
\label{eqn:bleu}
    BLEU = BP \times exp\left(\sum_{n=1}^{N}{w_n} log \left({p_n}\right)\right)
\end{equation}

\begin{equation}
\label{eqn:bleu-bp}
    BP =
    \begin{cases}
      1 & $$ c \geqslant r $$ \\
      exp\left(1-\frac{r}{c}\right) & $$ c < r $$\\
    \end{cases}  
\end{equation}

\noindent In Equation~\ref{eqn:bleu}, ${BP}$ refers to the brevity penalty. This variable aims to give a penalty to the generated sequence that is shorter than the ground truth. 
${r}$ and ${c}$ are correlated to the number of tokens in ground truth and candidate, respectively. 
Furthermore, ${p_n}$ is the \textcolor{black}{the modified precision for n-gram, ${w_n}$ are the weight and $\sum_{n=1}^{N}{w_n}=1$.}
 
\textcolor{black}{For the BLEU score, the higher it is, the better the result.}
In this work, we measure the performance of \method on the test data with cumulative n-gram BLEU score with n=4.
The cumulative score in the BLEU scores formula correlates to the individual n-gram scores at all orders from 1 to n. It then used the pre-defined weight to calculate the weighted geometric mean to obtain the final BLEU-n score.




\subsection{Research Questions}
\label{sec: rq}
In this work, we would like to investigate the following Research Questions (RQs):
\begin{itemize}
    \item{\textbf{RQ1:} \textit{Can we improve the performance of API sequence recommendation by leveraging information from \so for query expansion?}}
    
    While incorporating more information tends to generate a better result, it may not always be the case.
    In order to answer this RQ, we compare our baseline model with a model that is built with additional information from \so.
    We take CodeBERT with annotation input as the baseline and then experimented with \so title and \so API usage.
    The hypothesis for this question is that if we could find a semantically similar \so post for an annotation, information stored in the \so post would be imperative for the API sequence generation.
    
    In this work, we performed training on 2 types of data. One is where we only use annotation as the query to predict API sequence (i.e., the CodeBERT baseline), and the other one is where we incorporate \so posts information such as title and APIs as  query expansion (i.e., \method).
    Both models are then evaluated by comparing the BLEU score of the generated API sequence on the test set.    
    
    \item{\textbf{RQ2:} \textit{How should we utilize the information stored in \so post to improve API sequence generation performance?}}
    
    To answer this research question, we dig into the two types of \so information available in our \so dataset.
    Two components are being observed, namely the \so title and \so API.
    Note that we deliberately omitted \so body from this work due to the lengthy nature of the \so body.
    For the \so API, we utilize the extracted APIs mentioned in the accepted answer and treat them as related API tokens for a \so natural language (i.e., \so title).
    We explore the impact of these two components on the API sequence generation's performance.
    Our hypothesis for this question is that using more information from \so will accommodate a better API sequence generation.
    
\end{itemize}



\section{Result}
\label{sec:result}

In this section, we show the experimental results and answer the RQs.

\subsection{Answer to RQ1: }
\begin{table}[t]
    \caption {BLEU scores of \method compared to the baseline}
    \label{tab:res-rq1} 
    \centering
    \renewcommand{\arraystretch}{1.2}
    \begin{tabular}{@{}|l|l|l|l|l|@{}}
        \hline  
        \textbf{Approach} & \textbf{BLEU-1} &
        \textbf{BLEU-2} &
        \textbf{BLEU-3} &
        \textbf{BLEU-4} \\
        \hline
        \textbf{CodeBERT} & 0.49015 & 0.39946 & 0.33005 & 0.26296 \\
        \hline
        \textbf{\method} & \textbf{0.51371} & \textbf{0.42767} & \textbf{0.35959} & \textbf{0.29131} \\
        \hline
    \end{tabular}
\end{table}

As shown in Table \ref{tab:res-rq1}, in general, \method managed to outperform the vanilla CodeBERT by 10.8\% in terms of BLEU-4 score.
This shows that utilizing \so posts as query expansion that introduces additional information give better results than using only annotation in generating API sequence recommendation. The improvement is also consistent across the different BLEU scores.

\textcolor{black}{We also performed Mann-Whitney U hypothesis testing and got p-value less than 0.01, indicating that \method is statistically significantly better than the baseline.}




\begin{tcolorbox}[colback=gray!5!white,colframe=gray!75!black,boxrule=0.2mm]
\textbf{Answer to RQ1:} \method achieves the highest BLEU-4 score of 0.29131, which outperforms the baselines by CodeBERT by 10.8\%.
\end{tcolorbox}


\begin{table}[t]
    \caption {BLEU scores of different inputs}
    \label{tab:res-rq1-2} 
    \centering
    \begin{tabularx}{\linewidth}[t]{|X|X|X|X|X|}
        \hline
        \textbf{Input} & \textbf{BLEU-1} &  \textbf{BLEU-2} & \textbf{BLEU-3} &  \textbf{BLEU-4} \\
        \hline
        \textbf{annotation (baseline)} &  0.49015 & 0.39946 & 0.33005 & 0.26296\\
        \hline
        \textbf{annotation + SO title} & 0.50949 & 0.42230 & 0.35300 & 0.28507 \\
        \hline        
        \textbf{annotation + SO title + SO API} & \textbf{0.51371} & \textbf{0.42767} & \textbf{0.35959} & \textbf{0.29131}\\
        \hline
    \end{tabularx}
\end{table}

\subsection{Answer to RQ2: }

We performed an ablation study to measure the impact of leveraging \so title and \so APIs for query expansion.
Table~\ref{tab:res-rq1-2} displays the BLEU scores for each model that is built with different \so input(s) as the query.
It shows that leveraging \so information as query expansion can achieve a BLEU-4 score of up to 0.29131.
The highest BLEU-4 score is achieved by utilizing both \so title and \so API in the query.
Nonetheless, adding only \so title already improved the BLEU-4 score from 0.263 to 0.285, which translates to over 8\% improvement.
This shows that \so information brings a noticeable improvement when it is incorporated as query expansion.

\begin{tcolorbox}[colback=gray!5!white,colframe=gray!75!black,boxrule=0.2mm]
\textbf{Answer to RQ2:} Using \so title and \so API brings more improvement towards the API sequence generation technique compared to using only the \so title.
\end{tcolorbox}
\vspace{1mm}
\section{Discussion}
\label{sec:discussion}


\subsection{Ability to Find Similar \so Post}
\label{sec:discussion-similar-post}


We counted the number of cases in which the framework can successfully link an annotation to a relevant \so post.
We consider a \so post as relevant if the target API sequence shared common API(s) with API(s) mentioned in the \so answer post. 
We observed 3 degrees of success in finding a similar post, namely \textit{All match}, \textit{Partial match}, and \textit{No match}.
For the \textit{All match} category,
all of the APIs that are present in the target APIs also exist in the obtained in the answer post.
For the \textit{Partial match} category, a partial number of APIs are mentioned in the retrieved posts. 
The retrieved \so post contains one of the target APIs in its set of APIs. We categorized this kind of match as a partial match. 
As for the \textit{No match} category, none of the target APIs exist in the obtained answer post.
For example, we input a pair of annotation \textit{remove sub array} with target APIs of \texttt{Array.newInstance,System.arraycopy} and obtained a \so post with a title of \textit{Error making a sub-array}, which contains these APIs in the accepted answers: \texttt{System.arraycopy, Arrays.copyOfRange}.
The retrieved \so post contains \texttt{System.arrayCopy} method, which also appears in the target API.

\begin{figure}
\centering
\includegraphics[width=\linewidth]{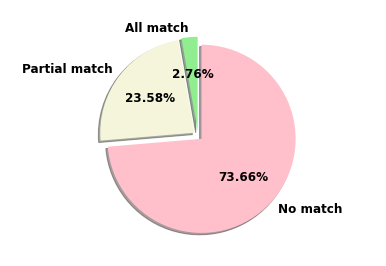}
    \caption{Capability of finding similar Stack Overflow post for the annotations in the training data}
    \label{fig:dicussion-a}
\end{figure}




Based on the overall result presented in Figure~\ref{fig:dicussion-a}, using the \so API in the post-processing phase, i.e., beam search, in order to limit the search space is not favorable in our case.
This statement is backed by our investigation, which shows that more than 73\% of the annotation in our dataset failed to find a relevant \so post that shares common API(s) with the target API.
Therefore, we deemed leveraging \so information in the query expansion manner is more suitable for our study because even though the API method(s) are not intersected, they may still come from the same class or domain.
For example, for annotation \textit{compress array} with a target API of \verb|ArrayList.size, Arraylist.ensureCapacity|, the retrieved \so post is a post with \textit{Expand array once it reaches limit} as its title, and the APIs mentioned in that post are \verb|Arrays.copyOf| and \verb|System.arraycopy|.

\begin{table}[t]
    \caption {Precision and Recall as Secondary Metric}
    \label{tab:rq1-precision-recall} 
    \centering
    \begin{tabularx}{0.7\linewidth}[t]{|X|X|X|X|X|}
        \hline
        \textbf{Input} & \textbf{Precision} &  \textbf{Recall} \\
        \hline
        \textbf{CodeBERT (baseline)} &  0.429 & 0.399 \\
        \hline
        \textbf{\method} & 0.452 & 0.442 \\
        \hline        
    \end{tabularx}
\end{table}

\textcolor{black}{Despite only about 26\% of the annotations being associated with a highly relevant \so post, \method was able to increase the performance of API sequence recommendation by nearly 11\%. This indicates that if similar \so posts were linked more effectively, the performance could be further enhanced.}

\textcolor{black}{
We also calculated precision and recall as additional metrics to supplement our analysis.
While precision and recall are commonly used for evaluating API recommendation techniques, they are more appropriate for assessing API set recommendations rather than API sequence recommendations.
Therefore, we primarily rely on the BLEU-4 score as the main metric in this study.
The purpose of calculating the precision and recall is to determine if \method can provide superior API recommendations or if it is only better at arranging the sequence of API calls compared to the baseline approach.
As presented in Table~\ref{tab:rq1-precision-recall}, \method achieves higher precision and recall scores than the baseline. 
The 10\% improvement in the precision indicates that \method manages to outperform the baseline in both accurately identifying the appropriate APIs and suggesting their correct invocation order.}





\subsection{Recommendation Analysis}
\label{sec:result-analysis}
In order to provide the future direction of API sequence generation, we discuss the cases where \method fails to provide an adequate sequence as well as the successful cases.
Based on our investigation, \method successfully incorporates \so information to improve the API sequence generation in 7,938 cases.
We also present several examples below to demonstrate how incorporating \so information guides the sequence-to-sequence model for predicting better API sequence.
Example 1 presents a case where \method could correctly identify the intention of the developer based on the annotation and managed to pair it with a \so post that improves the API sequence generation performance.

In Example 1, displayed in Figure~\ref{fig:eg1}, the annotation indicates that the developer wants to check whether a filename matches to a filter.
To do this, it could be done by using regex.
Without \so information, the baseline model predicts a completely irrelevant suggestion API sequence.
It could not be used to implement the intended functionality.
Nonetheless, introducing \so information to this datum successfully points the API sequence generation model towards the correct domain.
\method generated an API sequence that is relevant to the regex usage and is identical to the target API.

Similarly, in Example 2, the baseline produces a near identical API sequence to the target API. 
However, when \so information is introduced, \method manages to produced a sequence that is identical to what the developer has written.
It shows the capability of \method in generating API sequence with an assistance from \so information.



\lstdefinelanguage{qual}
{
  morekeywords={
    Annotation,
    Target_API,
    SO_title,
    SO_APIs,
    Baseline_prediction,
    PICASO_prediction,
  },
  sensitive=false, 
  morecomment=[l]{//}, 
  morecomment=[s]{/*}{*/}, 
  morestring=[b]" 
}

\begin{figure}[!htp]
\begin{center}
{
\lstset{language=qual,keywordstyle=\color{red}}
    \parbox{1\linewidth}{
\begin{lstlisting}[linewidth={\linewidth}, frame=tb,basicstyle=\scriptsize\ttfamily]
Annotation: 
check if the filename matches the filter

Target_API:
Pattern.compile Pattern.matcher Matcher.matches

SO_title: 
How to match letters only using java regex, matches method?

SO_APIs:
Pattern.compile, String.matches

Baseline_prediction:
String.lastIndexOf, String.length

PICASO_prediction:
Pattern.compile, Pattern.matcher, Matcher.matches
\end{lstlisting}
	}
}%
\caption{Example 1}
\label{fig:eg1}
\end{center}
\end{figure}

\begin{figure}[!htp]
\begin{center}
{
\lstset{language=qual,keywordstyle=\color{red}}
    \parbox{1\linewidth}{
\begin{lstlisting}[linewidth={\linewidth}, frame=tb,basicstyle=\scriptsize\ttfamily]
Annotation:
get the prefix of the object id

Target_API:
String.indexOf String.substring

SO_title:
Java regexp substring extraction pattern

SO_APIs:
String.indexOf, String.substring

Baseline_prediction:
String.lastIndexOf, String.substring

PICASO_prediction:
String.indexOf, String.substring
\end{lstlisting}
	}
}%
\caption{Example 2}
\label{fig:eg2}
\end{center}
\end{figure}

\begin{figure}[!htp]
\begin{center}
{
\lstset{language=qual,keywordstyle=\color{red}}
    \parbox{1\linewidth}{
\begin{lstlisting}[linewidth={\linewidth}, frame=tb,basicstyle=\scriptsize\ttfamily]
Annotation:
stop server

Target_API:
Thread.isAlive, Thread.interrupt, Thread.sleep, 
Socket.isclosed, Socket.close, ServerSocket.close

SO_title:
How exactly do I interrupt a thread?

SO_APIs:
Thread.currentThread, Thread.interrupted

Baseline_prediction:
ServerSocket.close, Thread.interrupt, ServerSocket.close

PICASO_prediction:
Thread.interrupt, Thread.sleep, Thread.currentThread, 
Socket.isClosed, Socket.close, ServerSocket.close
\end{lstlisting}
	}
}%
\caption{Example 3}
\label{fig:eg4}
\end{center}
\end{figure}

In Example 1 and 2, \method manages to produce a better API sequence recommendation than the baseline.
This is the advantage that is brought forward by the similar \so post.
The retrieved \so post contains title that is semantically relevant to the task described in the annotation, as well as informative \so API. 

Another evidence is served by Example 3, presented in Figure~\ref{fig:eg4}. Even though the annotation only consists of 2 words that convey an intention to stop a server, the SO title and the SO API provide an information on steps that one typically can do to stop a server. They point out that developers may need to interrupt a thread.
Leveraging this information, \method manages to perform well, generating an output that nearly identical to the target API.
On the other hand, we could see that even though the baseline contains 2 APIs from the target APIs, they are not in the correct order. It suggests the invocation of \texttt{ServerSocket.close} before invoking \texttt{Thread.interrupt}.
When \so information is introduced, the relative order between the APIs are correct. Specifically, \method manages to suggest the correct order of thread interruption by calling \texttt{Thread.interrupt, Thread.sleep, Socket.isclosed, Socket.close, ServerSocket.close}.


    








\begin{figure}[!htp]
\begin{center}
{
\lstset{language=qual,keywordstyle=\color{red}}
    \parbox{1\linewidth}{
\begin{lstlisting}[linewidth={\linewidth}, frame=tb,basicstyle=\scriptsize\ttfamily]
Annotation:
update all the necessary state to add node to slice

Target_API:
HashMap<Long>.put, HashMap<Slice,Long>.get,
HashMap<Slice,Long>.put, HashMap<Slice>.put


SO_title:
Updating mapped values for a 
TreeMap<String, TreeSet<String>>

SO_APIs:
Map.get, Map.put

Baseline_prediction:
ArrayList<Long>.add, ArrayList.add, ArrayList<Long>.add


PICASO_prediction:
HashMap<Long>.containsKey, HashMap<Long>.put,
HashMap<Slice,Long>.put, HashMap<Slice>.put
\end{lstlisting}
	}
}%
\caption{Example 4}
\label{fig:eg5}
\end{center}
\end{figure}

\textcolor{black}{
In Figure~\ref{fig:eg5}, we present Example 4 where \method produces a superior API sequence despite the relevant \so post not having any common APIs with the baseline. 
In this example, the developer intends to update the state of the nodes when they add a new node. As evidenced in the target sequence, the developer implements this function with a HashMap library. Interestingly, even though StackOverflow API does not share a common API, the title of the post (i.e., TreeMap) directs \method towards the usage of Map.
\method manages to generate a better API sequence recommendation for this case by suggesting the usage of HashMap’s API, instead of ArrayList, which is suggested by the baseline.
Therefore, we believe the improvement of the overall score comes from the cases where \so post guide \method towards better API sequence generation.
}

\begin{figure}[!htp]
\begin{center}
{
\lstset{language=qual,keywordstyle=\color{red}}
    \parbox{1\linewidth}{
\begin{lstlisting}[linewidth={\linewidth}, frame=tb,basicstyle=\scriptsize\ttfamily]
Annotation:
parse the line which holds the initial heap size

Target_API:
String.indexOf, String.substring, Integer.parseInt

SO_title:
Parsing string using the Scanner class

SO_APIs:
String.split, Integer.parseInt

Baseline_prediction:
String.indexOf, String.substring, Integer.parseInt

PICASO_prediction:
String.indexOf, Integer.parseInt
\end{lstlisting}
	}
}%
\caption{Example 5}
\label{fig:eg3}
\end{center}
\end{figure}

In Example 5, \method performs relatively worse than the baseline.
This is due to the missing API method invocation in the middle of the predicted sequence.
In the target API, \texttt{String.substring} is invoked before \texttt{Integer.parseInt}. 
In this case, the baseline predicts the sequence correctly.
However, when we introduced more information from \so, \method removes an API from the sequence, resulting in a drop of  BLEU-4 score.
This case suggests the importance of finding a highly relevant \so post.

All in all, in Examples 1-4, the retrieved \so contains SO title that are highly correlated to the annotation. This leads to the success of \method in producing a sequence that is closer to the target API. 
However, in Example 5, the retrieved \so post is less relevant as the SO title does not provide much additional information that can be used for query expansion. Rather, it may emphasizes some information over the other, which leads \method to generate a sequence that is further from the target API. 

\subsection{Lessons Learned}
\label{sec:lessons-learned}

    \textbf{Future studies should explore methods to better bridge the semantic gap between annotation and \so query.}
Even though sentence embedding can bridge the knowledge gap between Github and \so, i.e., code annotation and \so post, there is still a large room for improvement as 73.66\% cases are \textit{No match}. Thus, future research should explore better alternatives for doing it.
If we could retrieve a truly relevant \so post for a given annotation with a high success rate, we can also utilize the \so information beyond query expansion methods.
One option is to do a post-processing step in the generation process similar to Chen et al. work~\cite{chen2022more}.
\textcolor{black}
{
They proposed Cook, an API sequence recommendation technique that has enhanced with code-specific heuristic rules in the beam search module.
As they are working with partial source code as an input, they derived several code-specific heuristic rules, for instance, syntax-oriented heuristics.
Since we are working with annotation as an input and do not work with source code, we may not be able to derive a heuristic rule related to the syntax. However, it is possible to derive a rule about the sequence. 
For example, \texttt{Statement.close} method should be invoked after \texttt{Statement.executeUpdate} method.
If we can successfully retrieve relevant \so API for a given annotation, we could use it to limit the number of plausible APIs and use a heuristic rule to generate API sequence from them.}


\textbf{There exists more than one sequence to accommodate the developer's intention.} 
We need to find a way to objectively evaluate the API sequences generated by the Seq2Seq model from the perspective of usability, i.e., effectiveness and efficiency, i.e., length of API calls.
\textcolor{black}{
As discussed in Section~\ref{sec:result-analysis}, there exists more than one correct sequence to implement a functionality, as shown in Example 1.
There are also several functionalities that are known to have several implementations.
For example, official Java documentation page\cite{reading} published that there are at least 4 different ways to read a file in Java, each utilizing different API classes.
}
However, there are no suitable evaluation techniques to measure the effectiveness automatically other than manual evaluation.
One possible direction is to generate a dataset that contains test cases, so it could be used to judge whether the generated sequence could output a desirable result. It is a simple case though, since we cannot run the recommended API sequence directly to compare towards the test case.
A code generation method needs to be applied to craft a runnable code containing, transforming a sequence of API into a runnable snippet of code.


\subsection{Threats to Validity}
\label{sec:threat}

Threats to internal validity are associated with aspects of our experiment. It includes the replication of the baselines and the evaluation strategy.
The risk of mentioned threats was minimized by utilizing curated code and dataset from previous works.
For the dataset, we derived the dataset used in our experiment from the deduplicated dataset that is provided by Martin and Guo~\cite{martin2022deep}.
Furthermore, we also derived the sentence embedding training from CLEAR's replication package\footnote{\url{https://github.com/Moshiii/CLEAR-replication}}.
As for the baselines, we used the same code base for CodeBERT\footnote{\url{https://github.com/microsoft/CodeBERT}}.
Therefore, the risk of the contaminated dataset, incorrect re-implementation, and unfair evaluation should have been mitigated.


We consider the generalizability of our result as a threat to external validity.
In this study, due to the nature of the dataset from Martin and Guo~\cite{martin2022deep}, we only experiment with Java programming language and are limited to API that is present in the JDK library.
Moreover, we limited the API vocabulary to the APIs that are mentioned more than 5 times in the entire \so posts.

However, we believe that the impacts should be minimal 
as previous work on API recommendation \cite{deepapi}, \cite{wei2022clear}, \cite{huang2018api} also applied similar settings.
\section{Related Work}
\label{sec:related}

In this section, we review three lines of research that are most relevant to our work, i.e., (1) API recommendation, (2) mining API usage patterns, and (3) Mining Stack Overflow.


\subsection{API Recommendation}
In recent years, a multitude of API recommendation methods have been proposed~\cite{chen2021holistic,zhou2021boosting,rahman2016rack,deepapi}.
As mentioned in Section~\ref{sec:intro}, there are generally two types of methods for API recommendation.
We review several recent works in this part.

\textbf{\textit{Query-based} API recommendation:} 
Zhou et al.~\cite{zhou2021boosting} have proposed a framework BRAID (Boosting RecommendAtion with Implicit FeeDback) to boost the performance of query-based API recommendation systems.
BRAID adopts the user selection history as feedback information.
Moreover, it leverages learning-to-rank to re-rank the recommendation results.
Active learning techniques have been incorporated to alleviate the ``cold start'' of the limited feedback information at the beginning.
In addition, active learning can also speed up feedback learning.
The experimental results demonstrate that this framework can boost the performance of the three API recommendation systems.


\textbf{\textit{Code-based} API recommendation: }, they solve the task as the next-token prediction task. 
Xie et al.~\cite{xie2019hirec} propose HiRec, which is based on hierarchical context.
To generate hierarchical context, they use WALA~\cite{MainPage68:online} to generate call graphs of API methods and collect basic context from the call graphs.
They, we obtain a hierarchical structure of the surrounding project-specific code.
The final step is to combine both the basic context and hierarchical structure of surrounding project-specific code.
Another recent approach is APIRecX~\cite{kang2021apirecx}, which was proposed to handle the out-of-vocabulary issue of cross-library API recommendation.
It utilizes the GPT-based pre-trained subtoken language model.
At a high level, they use byte pair encoding~\cite{sennrich2016neural} to split each API call in each API sequence and pre-train a GPT-based language model.
By fine-tuning the pre-trained model, APIRecX can recommend APIs.
Different from them, our focus is to demonstrate that \so can be leveraged to boost query-based API recommendation.

\subsection{Mining API Usage Patterns}
Other than API recommendation, a similar long-researched topic is mining API usage patterns~\cite{xie2006mapo,zhong2009mapo,wang2013mining,fowkes2016parameter}.
MAPO is the first algorithm that has been proposed to mine API usage patterns from source code.
It was initially proposed by Xie and Pei~\cite{xie2006mapo} and further extended by Zhong et al.~\cite{zhong2009mapo}.
For a given query that describes the API, MAPO can leverage existing source code search engines to gather relevant source files and conduct data mining.
With the mined API usage patterns, MAPO has been extended to guide programmers in locating useful code snippets~\cite{zhong2009mapo}.
Wang et al.~\cite{wang2013mining} found that (1) there was a lack of metrics to measure the quality of mined API patterns, and (2) the API patterns mined by the prior approaches tend to be redundant.
Therefore, they focused on addressing these issues and proposed two quality metrics, i.e., succinctness and coverage, to measure the quality of mined API patterns.
Furthermore, they proposed an approach named Usage Pattern Miner (UP-Miner), which includes a two-step clustering strategy to mine succinct and high-coverage usage patterns of API methods from source code.
Similarly, Fowkes et al.~\cite{fowkes2016parameter} also identified the limitation of the prior approaches: the returned API calls tend to be large, redundant, and hard to understand.
To mitigate the issue, they propose PAM (Probabilistic API Miner), a near parameter-free probabilistic algorithm for mining the most interesting API call patterns. 

\subsection{Mining Stack Overflow}
As \so has been an essential part of software development, there is also rising research interest in mining \so posts.
Data from \so posts have been extensively used for multiple purposes.
Several studies have investigated leveraging \so as complementary of additional sources for specific tasks, such as augmenting API documentation~\cite{treude2016augmenting}, maintaining code~\cite{tang2021using}, and error fixing~\cite{wong2019syntax}.
Tang et al.~\cite{tang2021using} was the first to show that comment-edit pairs in \so can be potentially used for code maintenance.
They implemented an automated approach to link comments to code-snippet edits in \so.
Furthermore, they conducted a manual investigation of statistically representative random samples of the extracted comment-edit pairs.
They found that 50\% of the confirmed comment-edit pairs were general and related to correction, obsolete, flaw, and extension, which can be useful for general code maintenance tasks.
To demonstrate the potential use, they also leverage the confirmed comment-edit pairs to submit 15 pull requests to different GitHub repositories.
Among them, ten have been accepted.

Wong et al.~\cite{wong2019syntax} have focused on the syntax errors, and they extracted a Python dataset that contains human-made errors and their fixes from \so.
They first parse detected Python source snippet histories; second, they extract pairs of failed and fixed revisions; third, they validate pairs with interpreters; and finally, they record successfully evaluated pairs.
The resulting dataset is composed of real syntax errors made by developers, so it can potentially be used for the training and evaluation of code-related tasks, such as error detection.
By manual investigation, they found that errors made by \so users do not match errors made by student developers or random mutations.

Recently, Wu et al.~\cite{wu2023leveraging} focus on the task of identifying the relevant fragments of APIs.
They propose a new approach, SO2RT, to discover relevant tutorial fragments of APIs based on \so posts.
They utilize both the labeled information (relevance between \so Q\&A pairs and APIs) and unlabeled information (tutorial fragments and APIs).
They first automatically build two types of pairs, i.e., (1) relevant and irrelevant API and \so Q\&A pairs, and (2) API and tutorial fragment pairs.
They then train a semi-supervised transfer learning-based relevant fragment detection model, which aims to transfer the API usage knowledge in \so Q\&A pairs to tutorial fragments. 
Finally, the trained model can be used to infer relevant tutorial fragments of APIs.
\section{Conclusion and Future Work}
\label{sec:conclusion}
In this work, we propose an API sequence recommendation technique, \method which takes a natural language description of a programming task as input. 
It leverages a bi-encoder to conduct contrastive learning and a cross-encoder to build a classification model for finding the most semantically similar \so given an annotation (i.e., code comment).
\textcolor{black}{\method internally utilizes the title and APIs from similar \so post as a query expansion to fine-tune a pre-trained model resulting in a better API sequence generation model.}
Based on our experimental results, we show that the title and APIs are beneficial to boost the effectiveness of API sequence recommendation.
Specifically, \method improved approximately 11\% of the BLEU-4 score for the sequence recommendation.

In the future, we plan to investigate whether \so posts can help boost the performance of the API recommendation in other programming languages, such as Python.
We also plan to utilize the information from \so in other ways beyond query expansion. 
\textcolor{black}{Furthermore, it is also interesting to explore the usage of other parts of \so post such as its body to boost performance further. It may be beneficial as \so body contains detailed information about the post.}

\textcolor{black}{
Moreover, we plan to investigate how to use Stack Overflow posts to help with tasks beyond API recommendation, particularly those that involve natural language queries, such as bug report related tasks~\cite{wang2014version}. We also plan to investigate the challenges that developers face when using API recommendation tools, following prior works for other automated solutions, e.g., ~\cite{kochhar2015understanding, zou2018practitioners}.}

\textcolor{black}{\textbf{Availability.} Our replication package is publicly available at \url{https://github.com/soarsmu/Picaso}.}

\section*{Acknowledgement}
This research/project is supported by the Ministry of Education, Singapore, under its Academic Research Fund Tier 2 (Award No.: MOE2019-T2-1-193). Any opinions, findings, and conclusions or recommendations expressed in this material are those of the author(s) and do not reflect the view of Ministry of Education, Singapore.

\balance

\bibliographystyle{IEEEtran}
\bibliography{main}

\end{document}